\documentclass[twocolumn]{aastex631}
\usepackage{mathtools}
\usepackage{comment}
\usepackage{array}
\usepackage[utf8]{inputenc}
\usepackage{tabularx}
\usepackage{tabu}
\usepackage{multirow}
\shorttitle{Flare Following a Collision in a Galactic Nucleus}
\shortauthors{Brutman, Steinberg \& Balberg}

\newcommand{\MBH}{M_{\bullet}}
\newcommand{\Msun}{M_{\odot}}
\newcommand{\Rsun}{R_{\odot}}
\newcommand{\ergsmo}{\mathrm{erg\;s}^{-1}}

\published{\apjl\; OCT 2024}

\begin{document}
\title{The Primary Flare Following a Stellar Collision in a Galactic Nucleus}

\correspondingauthor{Shmuel Balberg}
\email{shmuel.balberg@mail.huji.ac.il}

\author{Yuval Brutman}
\affiliation{Racah Institute, Hebrew University of Jerusalem, Jerusalem, 91904, Israel}

\author[0000-0003-0053-0696]{Elad Steinberg}
\affiliation{Racah Institute, Hebrew University of Jerusalem, Jerusalem, 91904, Israel}

\author[0000-0002-0349-1875]{Shmuel Balberg}
\affiliation{Racah Institute, Hebrew University of Jerusalem, Jerusalem, 91904, Israel}

\begin{abstract}
High-velocity stellar collisions near supermassive black holes may result in a complete disruption of the stars. The initial disruption can have energies on par with supernovae and power a very fast transient. In this work we examine the primary flare that will follow the initial transient, which arises when streams of gas from the disrupted stars travel around the central black hole and collide with each other on the antipodal side with respect to the original collision. We present a simple analytic estimate for the properties of the flare, which depends on the distance of the collision from the central black hole and on the center of mass velocity of the colliding stars. We also present first of their kind radiation-hydrodynamics simulations of a few examples of stellar collisions and post-collision flow of the ejected gas, and calculate the expected bolometric light curves. We find that such post-collision flares are expected to be similar to flares which arise in tidal disruptions events of single stars.  
\end{abstract}


\section{Introduction\label{sec:Intro}}

Physical collisions between stars are inevitable in galactic nuclei, where stellar densities are high \citep{Roseetal2023}. Possible outcomes of stellar collisions include mergers, mass loss and mass transfer, and even partial or full destruction of stars \citep{Laietal1993,FreitagBenz2005,DaleDavies2006,RubinLoeb2011}. Obviously, the details depend on the relative velocity, the stars' masses and radii, their evolutionary state, and the impact parameter. 

A specific subclass of stellar collisions are such that occur close enough to the supermassive black hole (SMBH) which is commonly found in the centers of galactic nuclei \citep{KormendyHo2013}. Stars are in orbit around the SMBH, and their kinetic energies can easily exceed their individual binding energies, so that zero-to-low impact parameter collisions can disrupt the stars completely. \citet{BalbergSariLoeb2013} coined the term "collisional-supernovae" for such destructive collisions (DCs), since on the higher end of possible velocities, the total energy available can be on par with known stellar explosion mechanisms. The expected light curve arising from a DC of two main sequence stars was estimated by analytic approximations \citep{BalbergSariLoeb2013,HuLoeb2024a,AmaroSeoane2023}, and light curves from collisions between red giants were recently calculated with numerical simulations by \citet{Ryuetal2024} and \citet{Dessartetal2024}.

\citet{BalbergSariLoeb2013} also postulated that a DC and its light curve should be followed by a second, longer flare as the gas from the disrupted stars gradually approaches the SMBH and accretes onto it. This scenario is comparable to a tidal disruption event (TDE) of a single star by the SMBH. The general scheme in TDEs is that following the star's disruption, gas that remains bound to the SMBH proceeds on Keplerian paths, and eventually accumulates into an accretion disk (\citep{Rees1988}, see \citet{Gezari2021} for a review). Potential properties of such accretion flares following a DC were recently estimated analytically for the main sequence case by \cite{HuLoeb2024b}, while \citet{Ryuetal2024} included an analytical estimate for a spherical accretion variant following red giant collisions.

In this work we 
analyze the post DC flow and the flare 
that should be associated with it. We show that this flare is actually powered by shocks between streams in the flow, which converge to a collection of points that are antipodal to the location of the DC, as each mass element travels on its individual Keplerian path around the SMBH. The streams collide head-on with each other, dissipate their kinetic energy, giving rise to a bright flare which extends over time scales much longer than that of the initial DC. Such flares are comparable both in their typical locations and in their magnitude to those expected in collisions of streams from a single star disrupted in a TDE \citep{Piranetal2015,Shiokawaetal2015,Ryuetal2023,SteinbergStone2024}. 

The structure of the manuscript is as follows. In \S \ref{sec:simplemod} we discuss the physical picture for the emerging flow after a DC and describe a simple model for estimating the rate at which kinetic energy can be converted to an extended flare. In \S \ref{sec:simulations} we present first of their kind simulations of the post-collision flow, and calculate the resulting bolometric light curves. We summarize our conclusions and the potential observational implications in \S \ref{sec:conclusions}.   

\section{Prospects for a Post-collision Flare}\label{sec:simplemod}

Consider two Sun-like stars (each with mass $m_\star=\Msun$ and radius $R_\star = \Rsun$) colliding at a distance $r_0$ from an SMBH with mass $\MBH$. We assume that both stars have previously been captured in separate binary disruptions by the SMBH \citep{Hills1988}, and that both are near the periapse of eccentric orbits \citep{BalbergYassur2023}. Their velocities $\vec{v}_1$ and $\vec{v}_2$ then have magnitudes of about $v_{typ}=\sqrt{2G\MBH/r_0}$. The center of mass velocity of the two stars at their collision, $\vec{v}_{CM}$, can vary in magnitude between 0 and $v_{typ}$, and in direction with respect to the radius-vector from the SMBH, $\theta$. For clarity, we depict this radius vector from the SMBH to $r_0$ as the $+\hat{z}$ direction. Finally, the collision is also characterized by the impact parameter $b$ between the centers of the two stars. 

\subsection{A simplified model for the post collision flow and kinetic energy deposition}
\label{subsec:model}

We assess the initial conditions for the post-collision flow as follows. We assume that in the rest frame of the two-star center of mass all the gas which participated in the physical collision, $m_{ej}$, experiences full conversion of the available kinetic energy
\begin{equation}\label{eq:E_exp}
E_{exp}=\frac{1}{2}m_{ej} (v^2_{typ}-v^2_{CM})\;,
\end{equation}
to thermal energy. For an impact parameter $b=0$, $m_{ej}=2m_\star$. We further assume that the mass $m_{ej}$ then explodes spherically, where all the thermal energy is converted back to kinetic energy. Correspondingly, each mass element in this exploding mass, $dm$, is ejected at $r_0$ with a velocity $\vec{v}_0(dm)=\vec{v}_e(dm)+\vec{v}_{CM}$, where $\vec{v}_e$ is distributed spherically in the DC rest mass frame, and homologously in terms of the mass dependence $d|\vec{v}_e|/dm$ - assuming the ejecta is an exploding sphere with uniform density \citep{BalbergSariLoeb2013}. 

These assumptions define the distribution of velocities (magnitude and orientation) of the mass involved in the explosion following the two-star DC. We find that for most of the relevant ranges of the collision parameters, about one half of $m_{ej}$ (typically $46-54\%$, depending on $\vec{v}_{CM}$) is energetically bound to the SMBH. For each element of this bound mass, its initial velocity, $\vec{v}_0(dm)$, determines the Keplerian path the element will follow after leaving the explosion at $\vec{r}_0$. In particular, it is straight forward to calculate for each element its antipodal point "on the other side" of the SMBH, $r_{ap}(dm)$, and the time required for the mass element to reach that point, $t_{ap}(dm)$.  

Note that regardless of the magnitude and orientation of $v_{CM}$, any point in the bound mass has another point which is symmetric to it, in terms of the distance from the sphere's center and relative to the plane defined by the $z-$axis and $\vec{v}_{CM}$. This feature is depicted in figure \ref{fig:setup_figure} below. If $v_{CM}$ is oriented along the $z-$axis (including the case $v_{CM}=0$), there exists full azimuthal symmetry, and each point has an entire ring of such symmetric points. Hence, all mass elements in the bound mass have "partner-elements" which have the same $r_{ap}$ (on the negative $z-$axis) and $t_{ap}$. The partner elements will collide at the time $t_{ap}$ at their common $r_{ap}$, with velocities $v_{ap}(dm)$. 
\begin{figure}%[htbp] 
    \centering
    \includegraphics[width=\columnwidth]{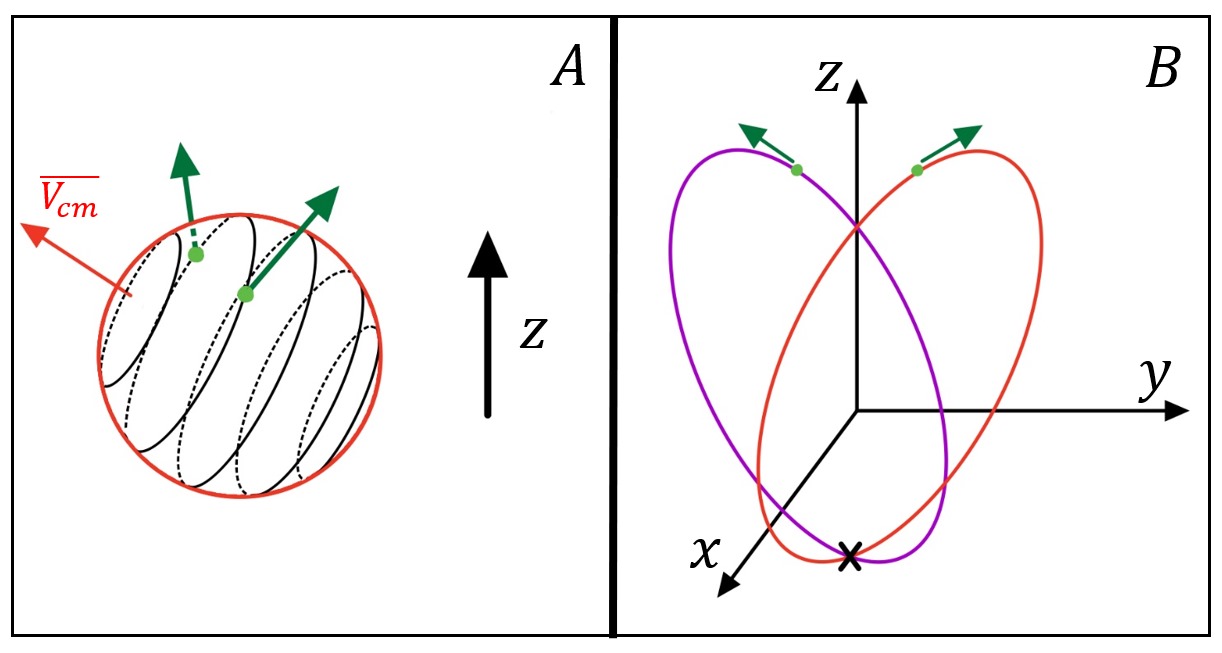} 
    \caption{The principal setup for the model of the ejecta following a stellar collision. The ejected mass is assumed to explode spherically and homologously in its rest frame (figure A), which has a velocity $\vec{v}_{cm}$ in the SMBH rest frame. There exists a symmetry with respect to the $\vec{v}_{cm}-\hat{z}$ plane for any two points on opposite sides of the plane defined by $\vec{v}_{cm}$ and the $\hat{z}$-axis. As a result, every mass element in the ejecta will collide with another element which is symmetric to it as their Keplerian paths coincide at the antipodal point to the collision (figure B).}
    \label{fig:setup_figure}
\end{figure}

The physical picture is therefore that all the bound mass collides at a collection of points on the $-z$-axis, and each element collides with its partner elements while carrying a specific kinetic energy of
\begin{equation}\label{eq:v_{ap}}
\begin{split}
E_{kin,0}(dm)= &\frac{1}{2}v^2_{ap}(dm)=\\
&\frac{1}{2} v^2_0(dm)+G\MBH\left(\frac{1}{r_{ap}(dm)}-\frac{1}{r_0}\right)\;.
\end{split}
\end{equation}
The stream-stream collisions run shocks through the gas, with a kinetic energy deposition rate, $\dot{Q}_{kin}(t)$, which is the cumulative sum over the rate of arrival of mass in the streams with $t_{ap}=t$, weighted by their appropriate values of $v^2_{ap}$.  

We suggest a minimal model for the primary flare: the stream-stream collisions mostly dissipate the kinetic energy to thermal energy, which is subsequently radiated away as a flare with an efficiency $\eta<1$. If this efficiency is constant, the resulting bolometric luminosity should track the temporal dependence of $\dot{Q}_{kin}(t)$. This is analogous to the standard model of TDEs which assumes that energy is radiated away as gas from the disrupted star falls back towards the SMBH. Note, however, that for TDEs this model suggests that kinetic energy is dissipated at a roughly fixed distance from the SMBH, so that the luminosity is proportional to the mass flow rate towards the SMBH, $\dot{M}(t)$.
In our model for the early flare we allow streams to collide over a wide range of $r_{ap}$, so we expect the light curve to follow the spatially integrated rate of dissipated kinetic energy. We also comment on the difference between these post-DC shocks and nozzle shocks expected in TDEs. While the latter also occur roughly at an antipodal point of the stream \citep{SteinbergStone2024}, they are a result of tidal compression along the normal to the orbital plane, with a compression velocity of order the escape speed of the unperturbed progenitor star. In our model here the shocks arise from head on collisions between streams at the orbital velocity.  

\subsection{Examples of Model Results}
\label{subsec:results}

For quantitative calculations of our model we used a simple Monte Carlo scheme for sampling the velocity distribution of the mass ejected in the explosion, according to the assumptions detailed above. By finding the appropriate values of $r_{ap}$ and $t_{ap}$ for each element of the bound mass, we generate a time dependent function of the kinetic energy source, $\dot{Q}_{kin}(t)$, of stream-stream collisions, as a function of the parameters of the two-star DC. We disregard  bound mass elements with $r_{ap}\leq 4R_S$ ($R_S$ is the SMBH Schwarzchild radius), assuming that such particles will be captured by the SMBH prior to completing their orbits \citep{Merritt2013}. 

\begin{figure}
    \centering
%    \begin{minipage}{\columnwidth}
        \centering
        \includegraphics[width=\columnwidth]{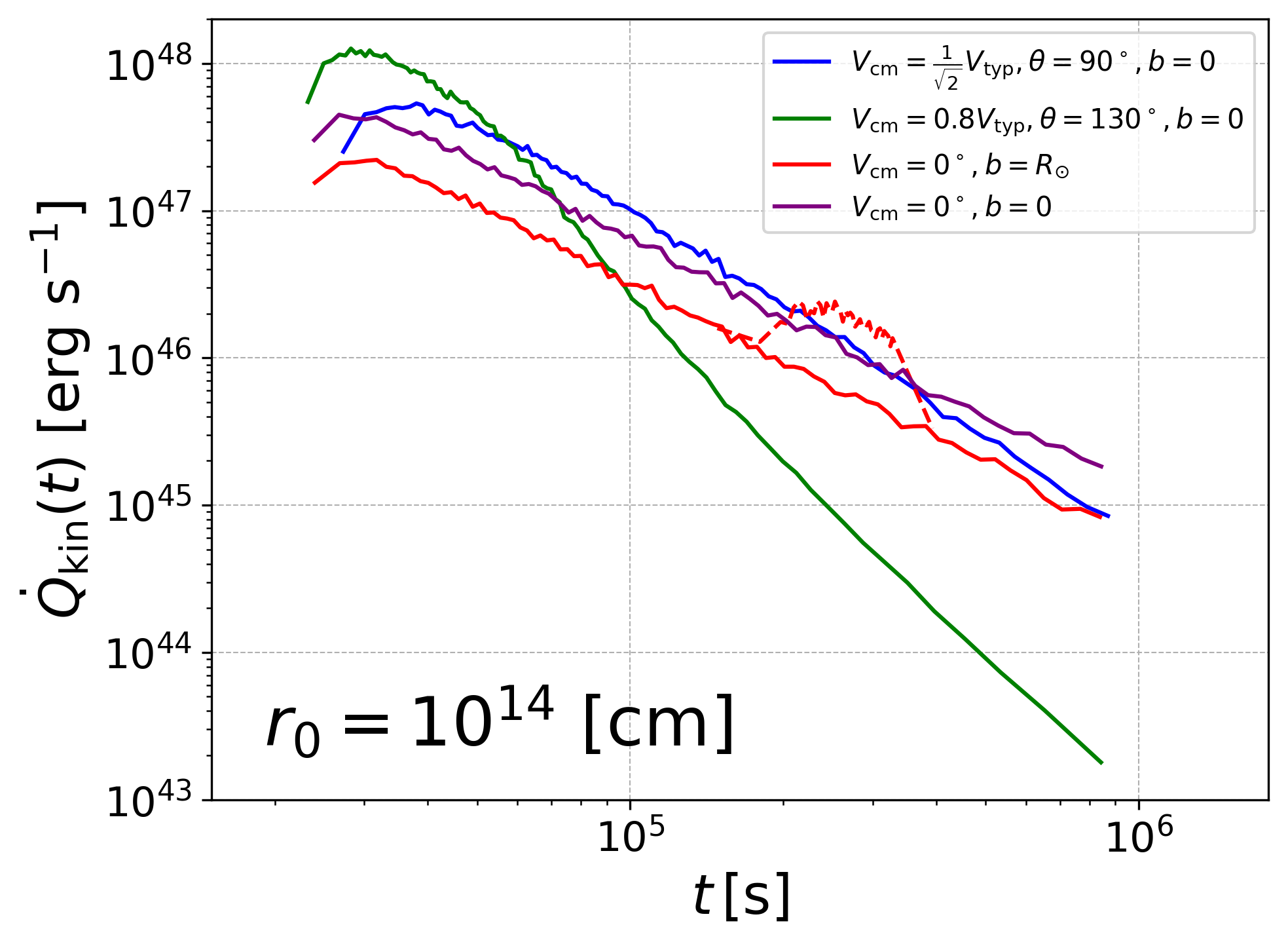} % Top image
        \includegraphics[width=\columnwidth]{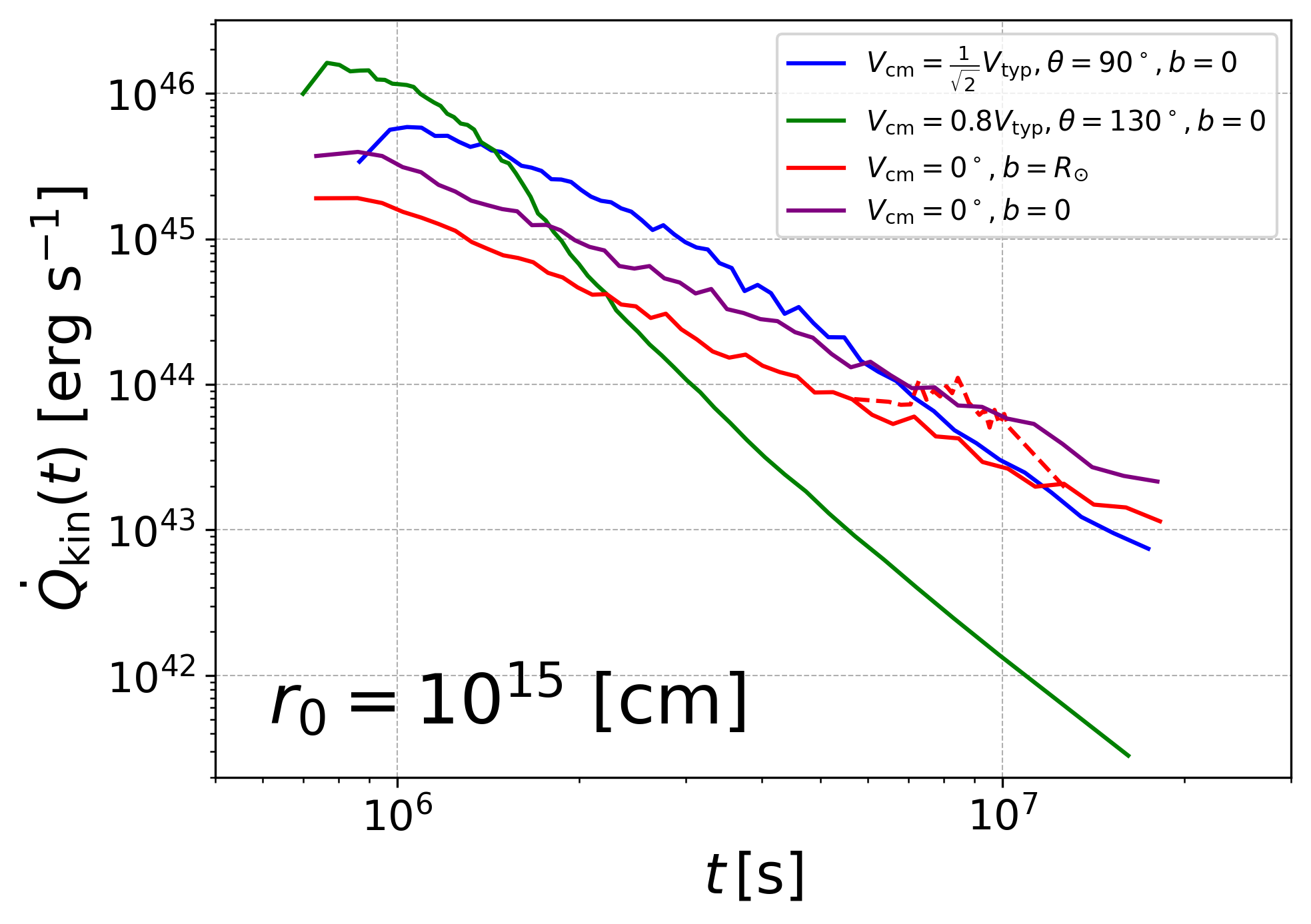} % Bottom image
        \caption{Kinetic energy deposition rates of gas streams colliding at the antipodal points with respect to the DC. The SMBH mass and stellar masses are $4\times 10^6\;\Msun$ and $1\;\Msun$, respectively, while the magnitude and orientation of $v_{CM}$ are varied. The DCs are set at $r_0=10^{14}\;\mathrm{cm}$ (top panel) or at $r_0=10^{15}\;\mathrm{cm}$ (bottom panel). The impact parameter in the DCs is set at $b=0$ or $b=\Rsun$; the dashed red curves display a crude model that includes mass which has not directly participated in the collision for a $b=\Rsun$ case (see text).}
        \label{fig:Ekindot}
%    \end{minipage}
\end{figure}

In figure \ref{fig:Ekindot} we demonstrate several such calculations of $\dot{Q}_{kin}(t)$. In all  calculations we fix the masses of the SMBH and stars at $\MBH=4\times 10^6\;\Msun$ and $m_\star=\Msun$, respectively, and consider DCs at $r_0=10^{14}\;\mathrm{cm}$ and $r_0=10^{15}\;\mathrm{cm}$. Each curve shows the consequences on $\dot{Q}_{kin}(t)$ of varying $\vec{v}_{CM}$ in orientation and magnitude. We mostly show calculations for a head-on collision with $b=0$, but also examine the case $b=\Rsun$.   

Several trends stand out from the results shown in figure \ref{fig:Ekindot}. First, by comparing the two panels we see, as is to be expected, that when the parameters of the DC are kept fixed, and only the radial distance $r_0$ is changed, the inflow of kinetic energy scales coherently. Time scales as $t_0\sim (r^3_0/G\MBH)^{1/2}$ and the magnitude of $\dot{Q}_{kin}(t)$ scales as $m_{ej}v^2_{typ}/t_0\propto r^{-5/2}_0$. 

\begin{figure}%[htbp]
    \centering
    \includegraphics[width=\columnwidth]{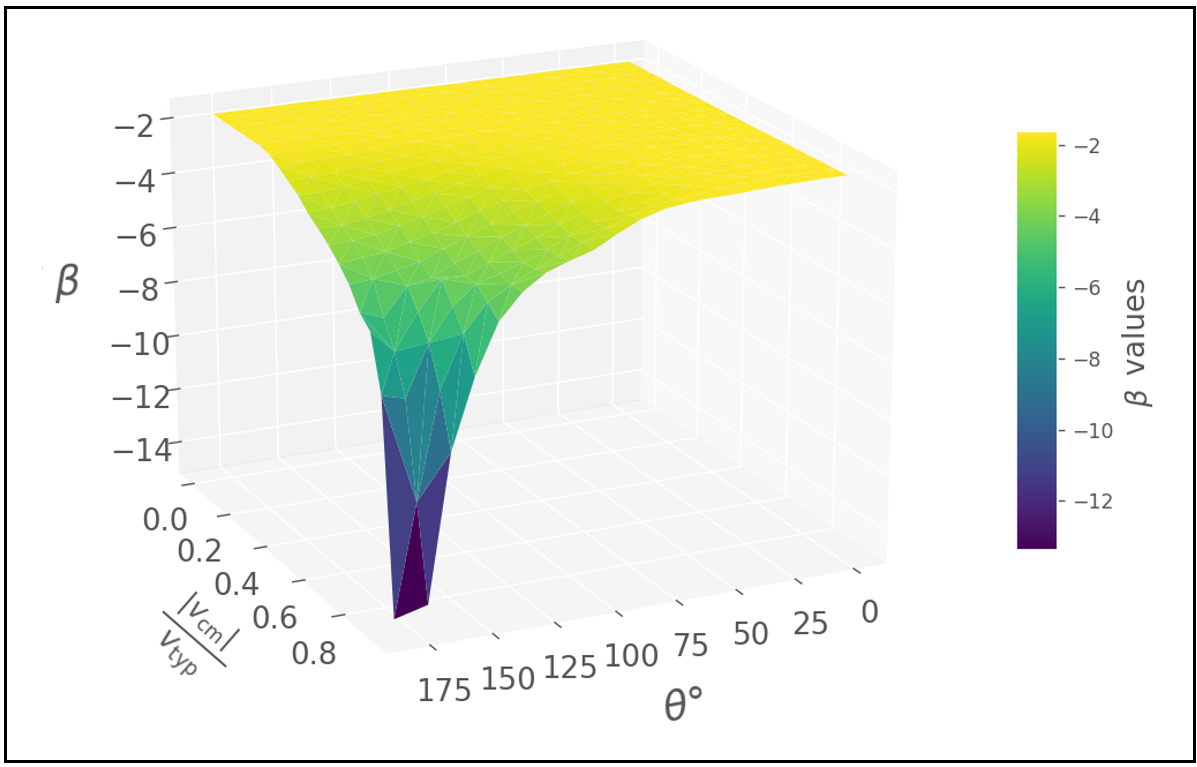} 
%    \captionsetup{font=footnotesize}
    \caption{A two-parameter survey of the late time power law dependence of $\beta\equiv d\ln\dot{Q}_{kin}/d\ln t$ assumed for the stream-stream collisions. The two parameters are $|\vec{v}_{CM}|$ (in units of $v_{typ}$) and the angle $\theta$. All calculations are for stellar collisions of two stars of mass $m=\Msun$ with $b=0$ at a distance of $r_0=10^{14}\;\mathrm{cm}$ from an $\MBH=4\times 10^6 \;\Msun$ SMBH.}
   \label{fig:beta}
\end{figure}

Secondly, changing the magnitude and orientation of $\vec{v}_{CM}$ creates observable differences between the energy deposition rates (compare different curves with $b=0$). In all cases $\dot{Q}_{kin}(t)$ peaks at $\sim t_0$ and falls of at later times, but the rate of the temporal decline can change significantly as a function of the parameters of the initial DC. This late time behaviour is observationally consequential, assuming that it is more likely to be observed than the relatively brief peak region. If the efficiency of conversion of the kinetic energy deposition to radiated luminosity is constant in time, we predict that the resulting light curve may be used to infer some of the properties of the DC. We expand on this point in figure \ref{fig:beta}, in which we plot the dependence of the late time power law decline rate, $\beta\equiv d \ln\dot{Q}(t)/d\ln t$, on the two parameters of the stellar collision, $|\vec{v}_{CM}|$ and $\theta$.     

We find that for most of the relevant range of $\{|\vec{v}_{CM}|,\theta\}$ combinations, $\dot{Q}_{kin}(t)$ tends to decline with a power law index of $\beta\gtrsim-2$. This value resembles the canonical power law index estimated for TDEs of $\beta\approx -5/3$ \citep{Phinney1989,LawSmithetal2020}. Consequently, this result suggests that observations of the late time declining part of the light curve following a DC near an SMBH may be difficult to distinguish from a TDE. Note, however, that in a small subsection of parameter space, with larger values of $|\vec{v}_{CM}|$ and $\theta$ (a DC when both stars are moving toward the SMBH) the energy deposition rates decline more steeply with time, with $\beta\leq -4$ (see the curves for $|\vec{v}_{CM}|=0.8v_{typ}$ and $\theta=130^\circ$ in figure \ref{fig:Ekindot}). Such sharply declining light curves, if observed, may serve as an indication of a flare following a DC

Finally, we consider the implications of a non-zero impact parameter. In the simple model described above, only material that has physically collided is assumed to contribute to $\dot{Q}_{kin}(t)$. For example, for an impact parameter of $b=R_\star$, only $\frac{1}{2}m_\star$ of each star participates in the collision ($m_{ej}=1m_\star$), and explodes with one half of the total kinetic energy held by the two stars. Correspondingly, the $\dot{Q}_{kin}$ curve in such a case simply scales by a factor of $\frac{1}{2}$ with respect to a $b=0$ collision with identical $\vec{v}_{cm}$ and $\theta$. This result is evident in the (solid red) curves for the case $\vec{v}_{CM}=0,\;\theta=0^\circ$ and $b=1\Rsun$ shown in figure \ref{fig:Ekindot}, when compared to the similar case but with $b=0$. 

It is implied above that material in segments of the stars which were beyond $b$ simply continues to travel with their initial velocity of $v\sim \sqrt{2G\MBH/r_0}$ (or $E_{tot}\approx 0$). Such material is not redirected towards the SMBH and does not contribute to the stream-stream collisions. 
However, the assumption that this material is unaffected by the collision is obviously crude. Some dissipation of kinetic energy is likely even in the extended parts of the stars. If so, some of this gas may also contribute to the stream-stream collisions. We demonstrate this directly in the numerical simulations shown below in \S \ref{sec:simulations}. Here we show a revised, fiducial model for such a case, in which for $\vec{v}_{CM}=0,\;\theta=0,\;b=1\Rsun$, we assume that for each star the material beyond $b$ in each star retains its original direction (orthogonal to the direction to the SMBH), but has a reduced specific kinetic energy, uniformly distributed between $0.5 v^2_{typ}/2$ and $0.75 v^2_{typ}/2$. The corresponding $\dot{Q}_{kin}(t)$ curve for this case is shown in figure \ref{fig:Ekindot} (dashed-red curves) as an addition to the initial result discussed in the previous paragraph. Since these two mass components have a narrower range of values for $r_{ap}$ and $t_{ap}$, they create a local maximum in the otherwise declining-in-time rate of the kinetic energy deposition rate.

Since at the high velocities typical of DCs gravitational focusing is inefficient, most collisions occur with a non-zero impact parameter. If so, primary flares following a DC should exhibit structure with a temporary rise in the light curve, instead of a monotonous decline. Improved models of this feature are clearly required, since if robust, it may distinguish DCs from TDEs.

\section{Radiation-Hydrodynamic Simulations}\label{sec:simulations}

We now turn to report the results of full radiation-hydrodynamic simulations of high velocity two-star collisions close to an SMBH. 
We simulate the collision between two solar mass stars using the Voronoi moving mesh Radiation Hydrodynamics code \textsc{rich} \citep{RICH, SteinbergStone2024}. Each star is modeled as an $n=1.5$ Lane-Emden polytrope using $10^6$ cells. The stars are given an initial velocity that corresponds to $\sqrt{2G\MBH/r_0}$, where we set $\MBH=4\cdot10^6\;\Msun$ and vary $r_0$ and the impact parameter. Table \ref{tab:runs} shows the initial conditions calculated in this work. 
In all of the runs, we assume a smoothed Paczyński–Wiita potential for the SMBH (as in \cite{SteinbergStone2024}) and ignore the self gravity of the gas. 
All our runs include dynamical radiation transport using gray flux limited diffusion, where the opacities are calculated using the atomic code \textsc{star} \citep{Star2, Star1}.

In order to better resolve the flow near the origin, we refine cells with a density above $\rho > 10^{-13}\Msun\Rsun^{-3}$ whose size is larger than $0.015\cdot r$. We also remove cells that are too small (size less than $0.6\;\Rsun$ or size less than $0.01\cdot r$ for cells within a distance of $10^3\Rsun$ from the origin).

\begin{table}[hbt]
\caption{Initial conditions for the radiation-hydrodynamic runs.}
\begin{center}
\begin{tabular}{ccc} 
\hline \hline
\multicolumn{3}{c}{List of Runs} \\
\hline
Impact parameter & $r_0$ & $\vec{v}_{cm}$ \\
\hline
0   & $10^{14}\;\mathrm{cm}$    & 0 \\
$\Rsun$ & $10^{14}\;\mathrm{cm}$ & 0 \\
$\Rsun$ &    $10^{14}\;\mathrm{cm}$ & $v_{typ}/\sqrt{2},\;\theta=90^{\circ}$ \\
0    & $10^{15}\;\mathrm{cm}$ & 0 \\
\hline
\end{tabular}
\end{center}\vspace*{-5mm}
\label{tab:runs}    
\end{table}

%\begin{table}%[hbt]
%\begin{center}
%\begin{tabular}{|p{1.5cm}|p{1.4cm}|p{2.5cm}|}
%\hline
%\multicolumn{3}{|c|}{List of Runs} \\
%\hline
%Impact parameter & $r_0$ & $\vec{v}_{cm}$ \\
%\hline
%0   & $10^{14}\;\mathrm{cm}$    & 0 \\
%$\Rsun$ & $10^{14}\;\mathrm{cm}$ & 0 \\
%$\Rsun$ &    $10^{14}\;\mathrm{cm}$ & $v_{typ}/\sqrt{2},\;\theta=90^{\circ}$ \\
%0    & $10^{15}\;\mathrm{cm}$ & 0 \\
%\hline
%\end{tabular}
%\end{center}
%\caption{Initial conditions for the radiation-hydrodynamic runs.}
%\label{tab:runs}
%\end{table}

\subsection{Dynamics}
\label{subsec:snapshots}
\begin{figure*}[t] 
    \centering
    \includegraphics[width=1\textwidth]
    {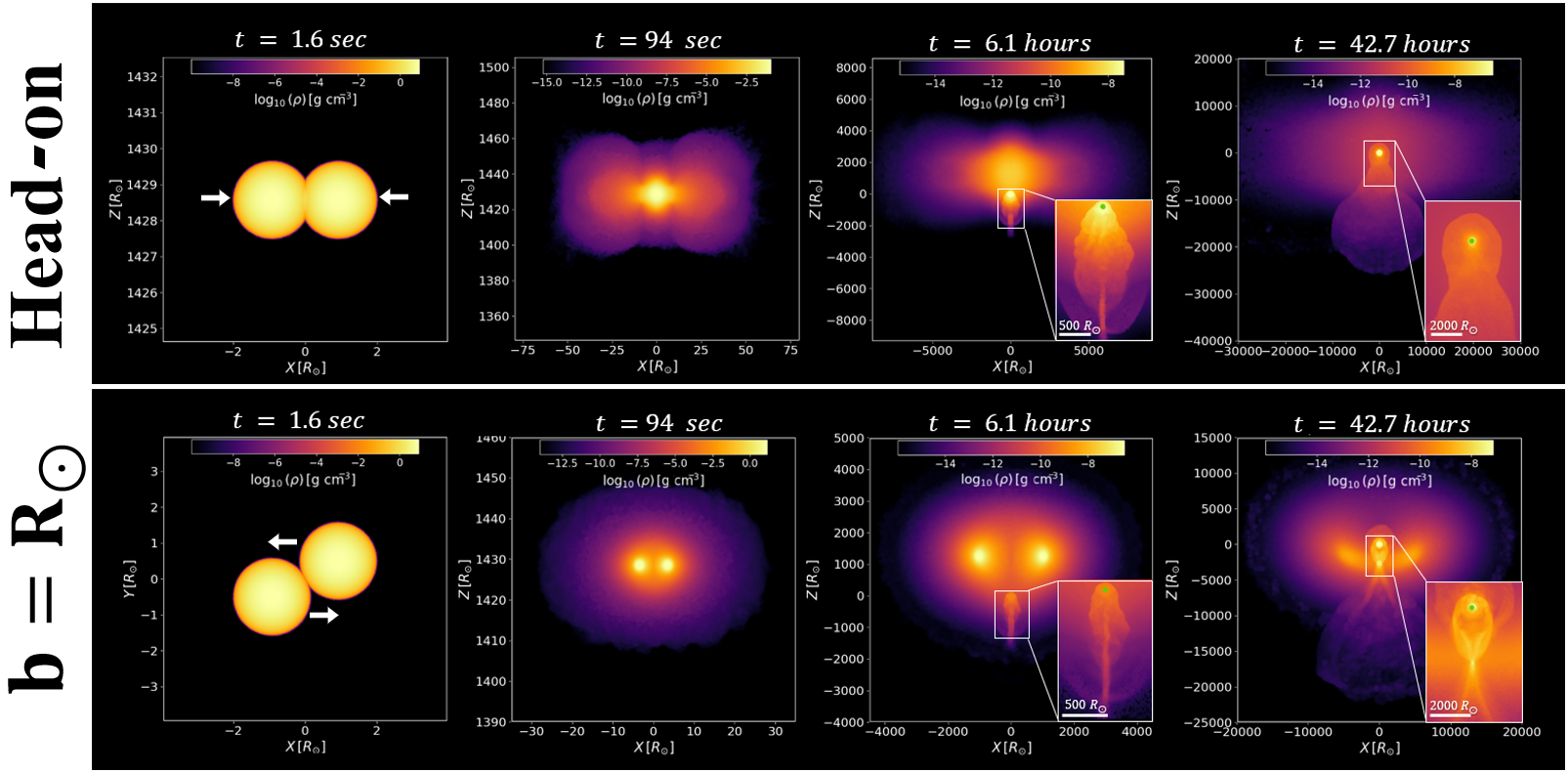} 
 %   \captionsetup{font=footnotesize}
    \caption{Snapshots with density maps of two RICH simulations. Top panel: a head-on DC between two Sun-like stars at a distance of $10^{14}\;\mathrm{cm}$ from an SMBH with mass of $\MBH=4\times 10^6 \Msun$. Bottom panel: a similar DC but the centers of the stars are shifted in the x-y plane by an impact parameter of $b=\Rsun$. In both simulations, the green dot in the extract box indicates the location of the SMBH (the origin).}
    \label{fig:snapshots}
\end{figure*}

Due to the large size required for the simulation box, we are limited in our ability to track the simulation over times much greater than $t_0$. Our main goal here is therefore to compare the flow calculated in the simulations to the simple model describe above and gauge the properties of the resulting light curves. We first focus on two simulations with DCs at $r_0=10^{14}\;\mathrm{cm}$: a head-on collision with $\vec{v}_{CM}=0$ and $b=0$, and another with $\vec{v}_{CM}=0$ but $b=\Rsun$, arranged by shifting the centers of the two stars in the $x-y$ plane. Figure \ref{fig:snapshots} shows four snapshots of these simulations ending at a time of $42.7\;$hours $(\sim 3.5t_0)$. 

The first snapshot of each simulation depicts the initial conditions of the DC. The second snapshot displays the outflow of the gas succeeding the disruption, at a time of $94\;$s after the collision ($\gtrsim 4 \Rsun/v_{typ}$). In the head-on collision the explosion is very radial, and also partially isotropic. Specifically, at time $t=94\;$s, we find that the mass averaged ratio of radial velocity to total velocity (in the collision rest frame) is $\sim 0.99$, and the ratio of mass ejected along the $\hat{z}-$axis to that ejected along the $\hat{x}-$ axis is $\sim 0.5$ (the two are arbitrarily defined by the masses with $v^2_z\geq 0.9v^2$ and $v^2_x\geq 0.9v^2$, respectively). These observations suggest that our simplified model for the kinetic energy deposition rate captures the qualitative nature of the post collision flow, and can serve as an quantitative approximation as well. 

The lower panel in figure \ref{fig:snapshots} shows that even in the $b=\Rsun$ case a significant fraction of the mass is also ejected over the entire $4\pi$ solid angle. However, there do exist large concentrations of mass which remain relatively compact, and move away from the center of the explosion; these are the mass segments of the two stars which did not physically participate in the collision, as discussed above. 

The third and forth snapshots in each simulation demonstrate the dynamics of the flow when streams of gas collide at their antipodal points (on the $-z$ side of the SMBH). The parts of the streams which arrive earliest are mostly composed of the mass elements with the smallest $r_{ap}$, and correspondingly carry higher specific kinetic energies. We find that as these streams collide, the high internal energies created by the shocks cause the gas to expand in a jet, in which most of the mass flows away from the SMBH (although some also flows toward it). As more material arrives in the streams, it accumulates at larger distances from the SMBH, creating an envelope through which the preexisting jet penetrates. It appears that the flow of the jet through the stalled gas creates a secondary, internal shock, which expands in the gas. This result is reminiscent of the "cocoon" shock structure which is often found in supernovae simulations when a fast jet travels through the stellar envelope (see, e.g., \citet{Gottliebetal2021} and references therein). 

The general trend at later times is that the shocked gas from the stream-stream collisions expands outward in what is a mostly radial flow, evolving as a plume around the $-\hat{z}$ direction. This occurs while more dilute gas in the streams continues to flow towards the locus of these collisions, generating ongoing shocks as they encounter the outflowing material. Interestingly, in the head-on DC case, at $t=42.7\;$hours the initial jet has already died out, and the entire flow appears to settle on a relatively smooth structure, with one extended shock layer. In the $b=\Rsun$ case, on the other hand, at this time there still exist denser substreams which are the remnants of the stellar segments that had not directly participated in the initial collision. As these substreams impact on the outflowing plume of gas, a more complex shock structure is created, including internal shocks which are not aligned with the $z-$axis. 

\subsection{Light Curves}
\label{subsec:lightcurves}
\begin{figure}%[htbp] 
    \centering
    \includegraphics[width=\columnwidth]{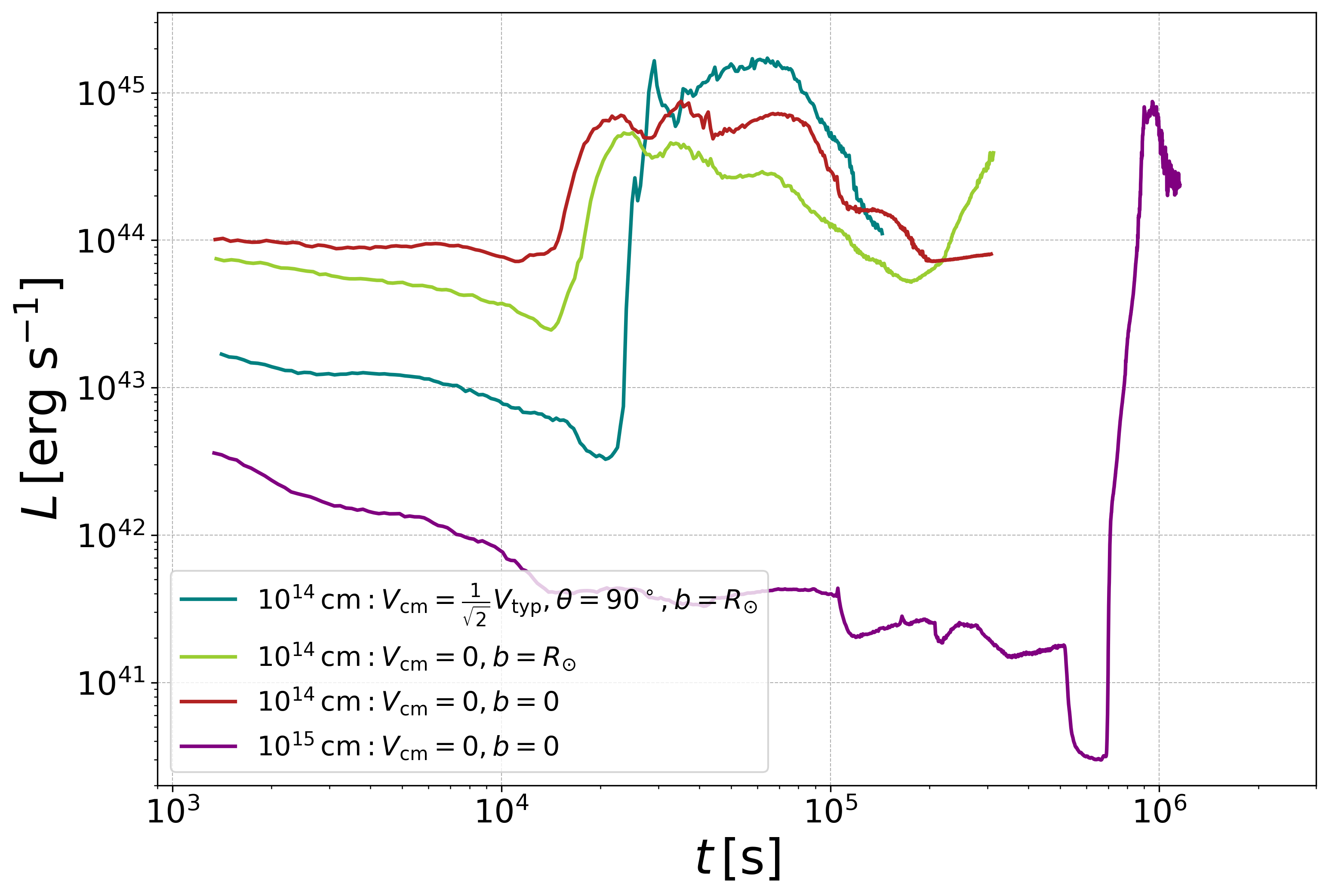} 
    \caption{Light curves for example cases, as noted in the legend. The red and green curves correspond to the scenarios shown in figure \ref{fig:snapshots}.}
    \label{fig:lightcurves}
\end{figure}

These qualitative and quantitative observations of the simulations carry over to the resulting light curves, calculated based on the flow density and temperature profiles. Figure \ref{fig:lightcurves} shows our calculated light curves corresponding to the two simulations described above. Our main interest here is in timescales of order $t_0$, and so for clarity we omitted the initial peak luminosity which results from shock breakout in the DC itself (and has typical timescale of $\sim \Rsun/v_{typ}$). We defer analysis of this shock breakout flash to subsequent work.    

The simulated light curves from the two simulations confirm some of the fundamental predictions of the simple model. Initially the luminosity is dominated by emission from the hot ejecta, and is roughly constant as this ejecta expands and cools (see \citet{BalbergSariLoeb2013} for a model of this phase). At a time of $\sim 2\times 10^4\;$s (about $t_0/2$) the luminosity rises sharply, which does indeed correspond to the first stream-stream collisions on the opposite side of the SMBH. Following this peak, the luminosity drops gradually, roughly as a $t^{\beta}$ power law with $\beta\gtrsim -2$, again as is expected in our simple model (see figures \ref{fig:Ekindot} and \ref{fig:beta}). We further note that the $b=\Rsun$ light curve roughly maintains a constant fraction of that of the $b=0$ case. This trend is consistent with our expectation that for a fixed $\vec{v}_{CM}$, the radiated power emitted in the stream-stream collisions tends to be proportional to $m_{ej}$. However, the simulated light curves do display some structure, as opposed to the smooth nature predicted in figure \ref{fig:Ekindot}. We suspect that this structure is due to the existence of internal shocks in the flow, as seen in figure \ref{fig:snapshots}. We specifically identify the significant rise of the luminosity in the $b=\Rsun$ case, just before the end of the simulation, with the arrival of the denser streams which are composed of mass components which did not collide physically. This is qualitatively consistent with the revised model discussed above, although the shape of the light curve is quantitatively different (compare to the dashed lines in figure \ref{fig:Ekindot}). 

Also shown in figure \ref{fig:lightcurves} is the bolometric light curve calculated for a simulation with $|\vec{v}_{CM}|=v_{typ}/\sqrt{2}$ and $\theta=90^\circ$ (the two stars were moving in orthogonal directions in the $x-y$ plane) and $b=\Rsun$. We again identify a general agreement with the simple model of \S \ref{sec:simplemod}. Specifically, the simulation recovers our expectation that in a $\vec{v}_{CM}\neq 0$ collision, the luminosity begins to rise at a similar time as for $\vec{v}_{CM}=0$, but continues to climb (gradually) to a higher peak luminosity, before it declines more steeply with time. Given the limited physical time over which we can track the flow, it is difficult to ascertain the long term power law of the light curve, but we do expect that on average, it should correspond to the prediction of the simple model, $\beta\approx -5$.

The bolometric luminosity estimated in the simulations with $r_0=10^{14}\;\mathrm{cm}$ peak around $10^{45}\;\ergsmo$, which corresponds to an efficiency of $\eta\approx 10^{-3}$ when compared to the available influx of kinetic energy, $\dot{Q}_{kin}$. This value implies that the vast majority of the kinetic energy dissipated in the stream-stream collisions is not radiated away, but rather reprocessed into kinetic energy as the shocked gas expands and cools. This is, of course, consistent with the general nature in shock driven explosions, such as supernovae. Since the gas is initially very optically thick (an optical depth of a few $10^{4}$ for a Thomson opacity), the internal energy is adiabatically converted to kinetic energy \citep{Arnett1996}. We note that this peak luminosity is of order the Eddington luminosity of the SMBH; while the overall flow is mostly not in an accretion pattern, it is possible that radiation transfer may generate some modulation on the dynamics at times close to the peak. Once the luminosity starts to decline, we expect that radiation transfer has a minimal effect on the flow.  

Interestingly, in a simulation with $\vec{v}_{CM}=0\;,b=0$ at $r_0=10^{15}\;\mathrm{cm}$, we find a peak luminosity due to the stream-stream collisions which is almost as high $(\sim 8\times 10^{44}\;\ergsmo)$.  When compared to the kinetic energy deposition rate, this implies a radiative efficiency of order $\eta\approx 0.1$. A higher efficiency for a larger $r_0$ is to be expected, since if the flow is entirely scalable, the optical depth at the region of stream-stream collisions should scale as $r^{-2}_0$. %This implies that for $r_0=10^{15}\;\mathrm{cm}$ the material in the collisions will need to expand only by a factor of a few to become optically transparent, and therefore release a larger fraction of its deposited internal energy. 
The surface area of the emitting region is also considerably larger, which may too contribute to an increased efficiency. 

A higher relative efficiency for larger $r_0$ may have significant observational consequences. If the magnitude of the bolometric luminosity is indeed relatively insensitive to the value of $r_0$, flares following a DC at larger distances from the SMBH may actually be more detectable, due to their longer times scales (in our simulations - months for $r_0=10^{15}\;\mathrm{cm}$ as opposed to days for $r_0=10^{14}\;\mathrm{cm}$). This point clearly requires further study, along with  the expected spectral features. We address these issues, along with extended simulation times, in future work.

\section{Conclusions and Discussion}
\label{sec:conclusions}

Galactic nuclei host a unique combination of an SMBH and a dense stellar cluster. One distinctive consequence of this environment is the potential for high velocity, destructive collisions between stars orbiting the SMBH. In this work we call attention to the fact that the primary flare from such a DC arises from shocks between streams of gas ejected from the destroyed stars. These streams travel in the gravitational field of the SMBH and converge to a locus of points which are anitpodal with respect to the location of the DC. The kinetic energy of the streams is converted to thermal energy, some of which is radiated away in what is a potentially observable light curve. 

We present a simple model for the flow of the ejected gas following the initial collision of the two stars, which allows to assess the rate at which kinetic energy in the streams is deposited in the region where the streams collide. The model includes an estimated dependence on the parameters of the stellar collision. 

We continue with full, three-dimensional radiation-hydrodynamic simulations of selected examples of the stellar collisions and the post collision flow in the gravitational field of an SMBH. The simulations suggest that the general properties of the post-collision flow are indeed consistent with the assumptions of the simple model, allowing to use it to infer the time scales, magnitude and time dependence of the stream-stream collisions, and especially the rate at which they dissipate kinetic energy. Moreover, the light curves calculated on the basis of the simulations indicate that these stream-stream collisions give rise to a bright flare, which extends over several dynamical times, $t_0=(r^3_0/G\MBH)^{1/2}$. This flare is therefore brighter and much longer than that produced due to the initial stellar collision, and appears to be the primary observational feature following a DC in a galactic nucleus.

We find that the bright flare peaks with a bolometric luminosity of $\sim 10^{45}\;\ergsmo$ around $t=t_0$, and then tends to gradually decline with time. Furthermore, the magnitude of the luminosity seems to be weakly dependent on $r_0$. We infer that as the typical volume in which the streams collide increases with $r_0$, so does the efficiency of radiating the produced thermal energy. For DCs which occur at smaller $r_0$ the available kinetic energy in the streams is higher, but the radiation efficiency appears to be significantly reduced. This result suggests that DCs farther from the SMBH may have a greater potential to be observed, since their primary flare extends over longer times. However, this preliminary finding requires further work, and especially extending our simulation times.

Generally speaking, it appears that the primary flare following a DC may appear similar in magnitude and temporal dependence to the flare following a TDE. This conclusion naturally raises the possibility that some observed TDE candidates may actually be results of DCs, rather than disruptions of single stars. A systematic investigation of this possibility is certainly warranted, and should involve a spectral analysis of the expected light curves. Extending the parameter survey carried out here is clearly necessary: in principle, there exists a larger parameter space for DCs than for TDEs, especially for non-zero impact parameter collision. In particular, DC flares could potentially create steeper or more structured light curves. These should be explored and compared to observed TDE candidates.  

\section*{Acknowledgements}
We recognize the support of the Interdisciplinary Computational Physics Laboratory (ICPL) at the Racah Institute of Physics for allocation of resources and technical assistance. We also thank A.~Loeb for helpful correspondence regarding the manuscript.

\bibliography{Flarebib}{}
\bibliographystyle{aasjournal}

\end{document}